\documentclass[a4paper,12pt,notitlepage]{article}

\usepackage[top=2.5cm, bottom=2.5cm, left=2.5cm, right=2.5cm]{geometry}
\usepackage{amsfonts}
\usepackage{amssymb}
\usepackage{amsmath}
\usepackage{graphicx}
\usepackage{pdfpages}
\usepackage{setspace}
\usepackage{multirow}
\usepackage[hidelinks]{hyperref}
\usepackage[title]{appendix}
\usepackage{algorithm}
\usepackage{algpseudocode}
\usepackage{tipa}
\usepackage{xcolor}

\usepackage[square,numbers]{natbib}
\makeatletter
\renewcommand{\fnum@figure}[1]{\textbf{\figurename~\thefigure}.}
\makeatother
\makeatletter
\renewcommand{\fnum@table}[1]{\textbf{\tablename~\thetable}.}
\makeatother

\begin{document}
\onehalfspacing

\begin{spacing}{1}
\title{\texttt{cissa()}: A \textsf{MATLAB} Function for Signal Extraction
\thanks{Financial support from the Spanish government, the contract grants MINECO/FEDER
ECO2016-76818-C3-3-P, PID2019-107161GB-C32 and PID2019-108079GB-C22/AEI/10.13039/501100011033 are
acknowledged.}
}
\author{
Juan Bógalo$^{a,}$
\thanks{Corresponding author. E-mail address: \texttt{juan.bogalo@telefonica.net} (J. Bógalo), \texttt{pilar.poncela@uam.es} (P. Poncela), \texttt{eva.senra@uah.es} (E. Senra).} , 
Pilar Poncela$^{a,b}$, 
Eva Senra$^{c}$\\ \\
\small{$^{a}$ \textit{Universidad Autónoma de Madrid, Spain}}\\
\small{$^{b}$ \textit{IBiDat, Univ. Carlos III de Madrid, Spain}}\\
\small{$^{c}$ \textit{Universidad de Alcalá, Alcalá de Henares, Spain}}\\
}
\date{\small{\today}}
\maketitle
\end{spacing}

\begin{abstract}
\texttt{cissa()} is a \textsf{MATLAB} function for signal extraction by Circulant Singular Spectrum Analysis, a procedure proposed in \cite{Bogalo_et_al:2021}. 
\texttt{cissa()} extracts the underlying signals in a time series identifying their frequency of oscillation in an automated way, by
just introducing the data and the window length.
This solution can be applied to stationary as well as to non-stationary and non-linear time series.
Additionally, in this paper, we solve some technical issues regarding the beginning and end of sample data points. 
We also introduce novel criteria in order to reconstruct the underlying signals grouping some of the extracted components.
The output of \texttt{cissa()} is the input of the function \texttt{group()} to reconstruct the desired signals by further grouping the extracted components.
\texttt{group()} allows a novel user to create standard signals by automated grouping options while an expert user can decide on the number of groups and their composition. 
To illustrate its versatility and performance in several fields we include 3 examples: an AM-FM synthetic signal, an example of the physical world given by a voiced speech signal and an economic time series.
Possible applications include de-noising, de-seasonalizing, de-trending and extracting business cycles, among others.

\bigskip


\end{abstract}

\newpage

\section{Introduction}
Signal extraction of the main features of a noisy time series is a key topic in many disciplines.
The interest in the extracted signals can range from just de-noising the time series, to de-seasonalizing or de-trending, to finding main oscillations or to the isolation of a group of frequencies of relevance to the analyst.
Expert analysis, as well as the automated and reliable processing of a large quantity of time series in real time is in high demand.
Within the physical world, signal processing is used in voice or pattern recognition.
In economics, Statistical Offices regularly produce de-seasonalized time series and estimations of business cycles widely used by analysts to assess, and forecast, the momentum of the economy.
Circulant Singular Spectrum Analysis, CiSSA, is a non-parametric signal extraction procedure, which is able to reconstruct the original time series as the sum of the orthogonal components of known frequencies.
CiSSA allows both, a detailed analysis of the frequency characterization of the data, as well as an automated application in those cases where the frequencies of interest are known before hand.

CiSSA, \cite{Bogalo_et_al:2021}, relies on Singular Spectrum Analysis (SSA), a widely used signal extraction alternative originating from the works of \cite{Broomhead_King:1986} and \cite{Fraedrich:1986}. 
At the same time, but in and independent way, research carried out in the former URSS proposed the 'Caterpillar' method, a variant of SSA, see \cite{Danilov_Zhigljavsky:1997}.
Different SSA variants work in two stages: decomposition and reconstruction. 
In the decomposition stage, the original time series data are transformed into a related trajectory matrix made of pieces of length $L$ (the window length) and perform its singular value decomposition obtaining the so-called elementary matrices.
In the reconstruction stage, the elementary matrices are grouped and time series recovered.
The original proposal is known as Basic SSA and the method used to perform the singular value decomposition introduces the different alternative versions of SSA. 
One of the alternatives, Toeplitz SSA, developed by \cite{Vautard_Ghil:1989}, computes the required eigenvectors by diagonalizing the matrix of second moments associated with the trajectory matrix. 
Traditionally, the identification of the frequencies associated with the elementary matrices is made after the time series have been reconstructed.
CiSSA, \cite{Bogalo_et_al:2021}, substitutes the variance-covariance matrix of the Toeplitz version by a related circulant matrix. 
CiSSA has an immediate advantage because the properties of circulant matrices allow for the exact identification between the eigenvectors and eigenvalues obtained in terms of the frequency of the components they represent.

Regarding the software to perform SSA, \cite{Golyandina_Korobeynikov:2014} propose an \textsf{R} package to compute Basic SSA in univariate time series, while \cite{Golyandina_et_al:2015} extend it to multivariate time series. Both contributions are based on the software of the Gistat Group, Ltd. (\url{www.gistatgroup.com}).

In this paper, we present several contributions not included in \cite{Bogalo_et_al:2021}. 
Firstly, we discuss the reconstruction of the underlying signals both at the beginning and the end of the sample proposing several alternatives that best suit different applications. 
Secondly, we introduce several grouping alternatives for different signal extraction problems that recognize automated, semi-automated and expert user options.
Finally, we provide two \textsf{MATLAB} functions to perform CiSSA: 
\texttt{cissa()} executes the algorithm after introducing the original data and the window length, $L$, producing as the outcome the exact decomposition into components of identified frequency and the power spectral density of the original time series; 
and \texttt{group()} allows the clustering of the identified components according to several degrees of automation that go from expert user to fully automated based on the needs of the analyst.

The structure of the paper is as follows: In section 2, we introduce the CiSSA algorithm. 
In section 3, we describe its extensions.
In section 4, we present the \texttt{cissa()} and \texttt{group()} \textsf{MATLAB} functions used to perform the algorithm and to form the desired reconstructed signals.
In section 5, we illustrate how CiSSA and the \textsf{MATLAB} functions work and show their performance in different fields and under alternative options.
Finally, in section 6, we present our conclusions.

\section{CiSSA algorithm}
CiSSA is an algorithm that exactly decomposes the original time series into the sum of a set of oscillatory components at known frequencies. 
Its main advantage is that, because the components are identified by frequency, the user can group them according to their needs. 
Grouping interests can be, for instance, in the form of previously determined frequencies, or as a dimension reduction technique paying attention to the selection of a certain amount of explained variability, among others.

In this section, we firstly present the main algorithm used to perform the frequency decomposition and the clustering procedures.

Let $\left\{ x_{t}\right\}$ be a zero-mean stochastic process, $t \in \cal T$ whose realization\footnote{For simplicity, we use the same notation for the stochastic process and for the observed time series. It will be clear from the context if we are referring to the population, or the sample. If it is not, we will explicitly clarify this in the main text.} of length $T$ is given by $\mathbf{x=(}x_{1},...,x_{T})^{\prime }$, where the prime denotes transpose, and let $L$ be a positive integer, called the
window length, such that $1<L<T/2$. 
The 2 stages in SSA (decomposition and reconstruction) are carried out in a four 4-step algorithm as follows:

\bigskip
\textbf{1st step. Embedding:}
We build a trajectory matrix by putting together lagged pieces of size $L$ from the original series.
The $L\times N$ trajectory matrix $\mathbf{X}$, $N=T-L+1$, is given by:

\begin{equation*}
\mathbf{X=}\left( 
\begin{array}{ccccc}
x_{1} & x_{2} & x_{3} & ... & x_{N} \\ 
x_{2} & x_{3} & x_{4} & ... & x_{N+1} \\ 
\vdots & \vdots & \vdots & \vdots & \vdots \\ 
x_{L} & x_{L+1} & x_{L+2} & ... & x_{T}%
\end{array}%
\right). \label{trajectory}
\end{equation*}

\bigskip
\textbf{2nd step. Decomposition:}
This step decomposes the trajectory matrix into elementary matrices of rank 1 that will be associated with different frequencies. In order to do so, we find the second order moments of the series, build a related circulant matrix $\mathbf{S}_{C}$ and calculate its eigenvalues and eigenvectors as functions of specific frequencies given by:
\begin{equation} 
w_k=\frac{k-1}{L}
\label{w_k}
\end{equation}
for $k=1,...,L$.

In order to do so, from the following estimated autocovariances
\begin{equation*} 
\widehat{\gamma }_{m}=\frac{1}{T-m}\sum_{t=1}^{T-m}x_{t}x_{t+m}, 
m=0,...,L-1,
\end{equation*} we compute the elements $\widehat{c}_{m}$ of the first row of a circulant matrix.
\begin{equation*}
\widehat{c}_{m}=\frac{L-m}{L}\widehat{\gamma }_{m}+\frac{m}{L}\widehat{\gamma }_{L-m},\qquad m=0,1,...,L-1\:. \label{c_m}
\end{equation*}

Therefore, we can build the circulant matrix given by:
\begin{equation*}
\mathbf{S}_{C}=\left( 
\begin{array}{ccccc}
\widehat{c}_{0} & \widehat{c}_{1} & \widehat{c}_{2} & ... & \widehat{c}_{L-1} \\ 
\widehat{c}_{L-1} & \widehat{c}_{0} & \widehat{c}_{1} & ... & \widehat{c}_{L-2} \\ 
\vdots & \vdots & \vdots & \vdots & \vdots \\ 
\widehat{c}_{1} & \widehat{c}_{2} & \widehat{c}_{3} & ... & \widehat{c}_{0}%
\end{array}%
\right) .
\end{equation*}

Based on \cite{Lancaster:1969}, the eigenvalues and eigenvectors of $\mathbf{S}_{C}$ for $k=1,...,L$ are given respectively by:
\begin{equation}
\widehat{\lambda} _{L,k}=\sum_{m=0}^{L-1}\widehat{c}_{m}\exp \left(\text{i}2\pi m\frac{k-1}{L}\right) \label{lambda}
\end{equation}

\begin{equation}
\mathbf{u}_{k}=L^{-1/2}(u_{k,1,}...,u_{k,L})^{H} \label{u_k}
\end{equation}
where $H$ indicates the conjugate transpose of a matrix and $u_{k,j}=\exp \left(-\text{i}2\pi (j-1)\frac{k-1}{L}\right) $.

Since the $k$-th eigenvalue in (\ref{lambda}) is an estimate of the spectral density at $w_k$ given by (\ref{w_k}), $k=1,...,L$, the $k$-th eigenvalue and corresponding eigenvector is associated with this frequency.

As a consequence, the diagonalization of $\mathbf{S}_{C}$ allows us to write $\mathbf{X}$ as sum of elementary matrices $\mathbf{X}_{k}$ as:
\begin{equation*}
\mathbf{X}=\sum_{k=1}^{L}\mathbf{X}_{k}=\sum_{k=1}^{L}\mathbf{u}_{k}\mathbf{w}_{k}^{\prime },
\end{equation*}
where $\mathbf{w}_{k}=\mathbf{X}^{\prime }\mathbf{u}_{k}$.

\bigskip
\textbf{3rd step. Basic frequency grouping:}
Given (\ref{w_k}) and the symmetry of the power spectral density, in this step, we group the elementary matrices by frequency.

The symmetry of the power spectral density leads to $\widehat{\lambda }_{k}$ $=$ $\widehat{\lambda }_{L+2-k}$, $k=2 \dots M$, where $M=\lfloor \frac{L+1}{2}\rfloor$.
The corresponding eigenvectors given by (\ref{u_k}) are complex, therefore, they are conjugated complex by pairs, $\mathbf{u}_{k}=
\mathbf{u}_{L+2-k}^{*}$ where $\mathbf{v}^{*}$ indicates the complex
conjugate of a vector $\mathbf{v}$. Then, $\mathbf{u}_{k}^{\prime }
\mathbf{X}$ and $\mathbf{u}_{L+2-k}^{\prime }\mathbf{X}$\
correspond to the same harmonic period. 

To obtain the elementary matrices by frequency, we first form the groups of 2 elements $
B_{k}=\{k,L+2-k\}$ for $k=2,...,M$ with $B_{1}=\{1\}$ and, $B_{\frac{L}{2}+1}=\left\{ \frac{L}{2}+1\right\} $ if $L$ is even. Secondly, we compute the
elementary matrix by frequency $\mathbf{X}_{B_{k}}$ as the sum of the two
elementary matrices $\mathbf{X}_{k}$ and $\mathbf{X}_{L+2-k}$, associated with
eigenvalues $\widehat{\lambda }_{k}$ and $\widehat{\lambda }_{L+2-k}$ \ and
frequency $w_k$ given by (\ref{w_k}), $k=1,...,M$,
\begin{eqnarray*}
\mathbf{X}_{B_{k}} &=&\mathbf{X}_{k}+\mathbf{X}_{L+2-k} \\
&=&\mathbf{u}_{k}\mathbf{u}_{k}^{\text{H}}\mathbf{X+u}_{L+2-k}
\mathbf{u}_{L+2-k}^{\text{H}}\mathbf{X} \\
&=&(\mathbf{u}_{k}\mathbf{u}_{k}^{\text{H}}+\mathbf{u}
_{k}^{*}\mathbf{u}_{k}^{\prime })\mathbf{X} \\
&=&2({\cal{R}}_{\mathbf{u}_{k}}{\cal{R}}_{\mathbf{u}_{k}}^{^{\prime }}+{\cal{I}}_{\mathbf{u}_{k}}{\cal{I}}_{
\mathbf{u}_{k}}^{^{\prime }})\mathbf{X}
\end{eqnarray*}
where ${\cal{R}}_{\mathbf{u}_{k}}$ denotes the real
part of $\mathbf{u}_{k}$ and ${\cal{I}}_{\mathbf{u}_{k}}$ its imaginary part.
Notice that the matrices $\mathbf{X}_{B_{k}},k=1,...,M,$ are real. 

As a by-product, the algorithm also allows us to determine the share of explained variability at frequency $w_k$ given by $sh_k= 2\times \widehat{\lambda }_{k}/\sum\widehat{\lambda }_{k}$ for $k=2,...,M$ and $sh_1=\widehat{\lambda }_{1}/\sum\widehat{\lambda }_{k}$ for $k=1$.

\bigskip
\textbf{4th step. Reconstruction:}
Finally, this step transforms the matrices obtained in step 3 into $M$ signals of the same length as the original series for frequencies $w_k$ given by (\ref{w_k}), $k=1,...,M$, denominated elementary reconstructed components by frequency.

To transform $\mathbf{X}_{B_{k}} = \left(\widetilde{x}_{i,j}\right)$, $k=1,...,M$, into a time series $\widetilde{x}_{t}^{(B_{k})}$, we use the algoritm called diagonal averaging derived by \cite{Vautard_et_al:1992}. This consists of averaging the elements of $\mathbf{X}_{B_{k}}$ over its antidiagonals, i.e., the hankelization of this matrix with the operator $\text{H}\left(\cdot\right)$:

\begin{equation*}
\widetilde{x}_{t}^{(B_{k})}=\text{H}\left(\mathbf{X}_{B_{k}}\right) = \left\{ 
\begin{array}{l}
\frac{1}{t}\sum_{i=1}^{t}\widetilde{x}_{i,t-i+1},\qquad 1\leq t<L \\ 
\frac{1}{L}\sum_{i=1}^{L}\widetilde{x}_{i,t-i+1},\qquad L\leq t\leq N \\ 
\frac{1}{T-t+1}\sum_{i=L-N+1}^{T-N+1}\widetilde{x}_{i,t-i+1},\qquad N<t\leq T.
\end{array}
\right .
\end{equation*}

\bigskip
We have created \texttt{cissa()}, a \textsf{MATLAB} function, to implement the CiSSA algorithm that provides the $M$ elementary reconstructed components by frequency and the power spectral density of the original time series.
We will illustrate how it works in section 4.

\section{Extensions in CiSSA}
In this section, we introduce new results with respect to \cite{Bogalo_et_al:2021} that allow us to answer two technical issues when implementing the algorithm. 
The first is related to the end of sample data points, a topic usually discussed in other signal extraction techniques but not in SSA until now.
The second deals with further grouping of the estimated components in different contexts. 
This second issue has been approached in SSA in different ways and is still an open topic within this methodology. 
Because one of the advantages of CiSSA relies on matching principal components with frequencies, we can use this result to form groups with the desired frequencies. 
Therefore, CiSSA can be used to extract signals at desired frequencies, i.e., acting as a band-pass filter. 
We have also introduced some new alternatives in order to form the groups, either in an expert user, semi-automated or fully automated way.

\subsection{Beginning and end of sample data points}
Notice that the number of times that a data point appears in the trajectory matrix depends on its proximity to the left upper, and right lower, corners of the matrix. 
Therefore, in step 4 of the algorithm, reconstruction, we may use a different number of elements when reconstructing each point of the unobserved extracted signals as we move to the two aforementioned corners of the elementary matrices by frequency. 
These two corners represent the two extremes of the data, the beginning and the end of the sample. 
As a consequence, the reconstruction at the extremes can be rather unstable because it is done with very few elements. 
Notice that, for both extremes $t=1$ and $t=T$, it is done with only one element. 
To overcome this problem, we propose two alternative solutions depending on the type of signals. 
The \textsf{MATLAB} function, \texttt{extend()}, gives the possibility of implementing these different alternatives through a parameter that can be specified by the user as an optional input to \texttt{cissa()}.

\bigskip
{\bf First solution: autoregressive forecasting/backcasting.}
This consists of adjusting an autoregressive (AR) process to the first difference of the time series. The order of the AR process is large enough to pick up the dynamics of signals of very different nature. 
In particular, we use $p=T/3$ with $p$ being the order of the AR process. 
We can make forecasts with this auxiliary AR process and afterwards integrate them in order to recover the level of the observed data. 
The process is similar at the beginning of the sample, but backcasts are used instead of forecasts. 
The number of forecasts used in either extreme is the window length $L$.
This option is of general use for any time series. 

\bigskip
{\bf Second solution: Mirroring.} 
This proposal goes in line with that of \cite{Gianfelici_et_al:2005} and consists of mirroring the time series around the extremes. 
In this way, after we reach the end of the series at $T$, we add data points considering $x_{T+j}=x_{T-j+1}$, $j=1,...,T$. 
A similar solution is applied at the beginning of the sample with $x_{1-j}=x_j$, $j=1,..,T$.
This extension can be used for stationary time series, and also works well for certain types of non-stationary time series such as amplitude and frequency modulated (AM-FM) signals but does not necessarily work well for trending time series.

\subsection{Grouping}
The outcome of the automated CiSSA algorithm is a set of $M$ time series, one for each frequency, $w_k$ given by (\ref{w_k}), $k=1, \cdots, M$. 
Because the components are already identified by frequency, afterwards the user can group them according to their needs. 
The \textsf{MATLAB} function, \texttt{group()}, gives the possibility of implementing different grouping strategies after executing \texttt{cissa()}. 
Alternatives range from a completely expert user option defining the groups after analyzing the data, incorporating semi-automated options where the user introduces simply a parameter related to the amount of variability they want to explain, or the relevance of the \texttt{psd} values they want to introduce, and even to providing a totally automated option devoted to economic time series.

\bigskip
{\bf Users in Economics.} Under this option economists only need to introduce the number of observations per year and automatically obtain the default signals typically required in a time series of economic activity: trend, business cycle and seasonality. 

\bigskip
{\bf Expert user.} With this option the user introduces their own groups according to their previous knowledge of the problem or attending to their analysis of the power spectral density obtained in \texttt{psd} as an output from \texttt{cissa()}.

\bigskip
{\bf Dimension reduction.} 
In this case, the \texttt{group()} function retains as many components as necessary to achieve a pre-specified amount of explained variability. 
It produces a single reconstructed time series as the sum of these components. Furthermore, this function also provides the selected frequencies. 

\bigskip
{\bf Largest \texttt{psd} grouping.}
Now, the \texttt{group()} function considers the empirical distribution of the eigenvalues and selects those over a percentile specified by the user. 
The function, again, produces a single reconstructed series as the sum of these components and provides the selected frequencies and the amount of variability explained.

\section{MATLAB functions}
There are two main functions available to the user: \texttt{cissa()} and \texttt{group()}. 
The \texttt{group()} function needs to be run after \texttt{cissa()} and cannot be run on its own because it uses as inputs the output of \texttt{cissa()}. 
Firstly, we describe \texttt{cissa()} below.

\subsection{The cissa function} 
The \texttt{cissa()} function performs CiSSA. 
Given an input series $x_t$ and a window length, $L$, it returns the $M$ elementary reconstructed components $\widetilde{x}_{t}^{(B_{k})}$ by frequency given by (\ref{w_k}), $k=1,...,M$, and the estimated power spectral density of the input time series. The syntax of the function is:

\bigskip
\begin{ttfamily}
\begin{slshape}
\noindent
[Z, psd] = cissa(x,L,H)
\end{slshape}
\end{ttfamily}
\bigskip

\noindent
where the inputs of the function are:
\begin{itemize}
\item \texttt{x}: Columm vector containing the original data.
\item \texttt{L}: Window length.
\item \texttt{H}: Optional. A parameter related to the characteristics of the time series and the kind of extension made at the beginning and end of the sample. It can take the following values:
\begin{itemize}
\item  \texttt{H=0} (Default) Autoregressive extension.
\item  \texttt{H=1} Mirroring.
\item  \texttt{H=2} No extension.
\end{itemize}
\end{itemize}
The parameter \texttt{H} is the input of the internal \texttt{extend()} function. 
The option \texttt{H=0} extends the data using an autoregressive forecasting/backcasting. 
It is indicated for any time series. 
The option \texttt{H=1} corresponds to mirroring the time series.
It can be used with stationary time series and it also works well for AM-FM signals.
Finally, the option \texttt{H=2} does not extend the data. 
It is suitable for deterministic time series.
After performing CiSSA, the function returns the following outputs:
\begin{itemize}
\item \texttt{Z}: Matrix whose $M$ columns are the elementary reconstructed components by frequency.
\item \texttt{psd}: Column vector with the estimated power spectral density estimated at frequencies $w_k$, $k=1,...,L$ through the eigenvalues of the circulant matrix of second moments.
\end{itemize}

\subsection{The group function}
Once we have extracted the components (saved in \texttt{Z}) with \texttt{cissa()}, we can use the \texttt{group()} function to reconstruct the desired signals. 
The second \textsf{MATLAB} function that we provide uses as input arguments the two outputs from \texttt{cissa()} and an additional argument to specify how we want to perform the grouping. 
All in all, this \textsf{MATLAB} function groups the reconstructed components by frequency obtained with CiSSA into disjoint subsets and computes their share of the corresponding power spectral density. The syntax of the function is:

\bigskip
\begin{ttfamily}
\begin{slshape}
\noindent
[rc, sh, kg] = group(Z,psd,I)
\end{slshape}
\end{ttfamily}
\bigskip

\noindent
where the inputs of the function are:
\begin{itemize}
\item \texttt{Z}: Matrix whose $M$ columns are the elementary reconstructed components by frequency obtained with \texttt{cissa()}.
\item \texttt{psd}: Column vector with the power spectral density estimated at frequencies $w_k , k=1,2,...,L$, given by (\ref{w_k}) obtained with \texttt{cissa()}.
\item \texttt{I}: This input argument has four different options:
\begin{enumerate}
\item A positive integer. It is the number of data per year in an economic time series. The function automatically computes the trend (oscillations with period greater than 8 years), the business cycle (oscillations with period between 8 and 1.5 years) and a seasonal component.
\item A cell array. Each cell contains a row vector with the desired values of $k$ to be included in a group, $k=1,2,...,M$. The function computes the reconstructed components for these groups.
\item A number in the interval (0, 1). This number represents the accumulated share of the psd achieved with the sum of the share associated with the largest eigenvalues. The function computes the original reconstructed time series as the sum of these components.
\item A number in the interval (-1, 0). 
This number is a percentile (in positive) of the empirical distribution of the values of \texttt{psd}. 
The function computes the original reconstructed time series as the sum of the reconstructed components by a frequency whose \texttt{psd} is greater than this percentile.
\end{enumerate}
\end{itemize}
After running this function, we obtain as output the following items:
\begin{itemize}
\item \texttt{rc}: Matrix whose columns are the reconstructed components for each group or subset of frequencies. 
In the case of economic time series this matrix has 3 columns corresponding to the trend, business cycle and seasonal components, respectively. 
Under option 2 (cell array), it returns as many columns as groups introduced.
And under options 3 and 4 (largest \texttt{psd} and dimension reduction respectively), this matrix has just one column. 
\item \texttt{sh}: Column vector with the share of the power spectral density for each group.
\item \texttt{kg}: Cell array where each cell contains a row vector with the values of $k$ belonging to a group. Option 1) produces 3 groups, option 2) gives the groups introduced by the user and options 3) and 4) produce a single group. In option 3), the values of $k$ are sorted according to the share of total psd of their corresponding eigenvalues.
\end{itemize}

\section{Applications}
In this section we illustrate the use of the functions \texttt{cissa()} and \texttt{group()} through three examples.
Firstly, we will consider a case where the frequencies of oscillation of the underlying components, are known to the analyst. This is an AM-FM synthetic signal and shows the performance of CiSSA in an expert user way. 
Secondly, we will analyze a physical signal from voiced speech recognition illustrating how \texttt{cissa()} can process these type of signals. Within this example, we illustrate the use of options 2, 3 and 4 of the function \texttt{group()} through the analysis of the voiced signal of the word "Hello". Finally, we present an economic example to show the performance of \texttt{cissa()} and \texttt{group()} in an automated way.

\subsection{Synthetic Signal}
We have simulated a synthetic signal to illustrate the expert user alternative to define the groups in the \texttt{group()} function.
The original signal is the sum of two components: the first is an AM signal and the second is both an amplitude and frequency modulated (AM-FM) signal, similar to that used in \cite{Biagetti_et_al:2015}. 
Let $x(t)=x_1(t)+x_2(t)$ where $x_1(t) = a_1(t) cos(w_1t)$ and $x_2(t) = a_2(t) sin(w_{2,0}t + w_{2,1}\frac{t^2}{2T})$ with $a_1(t) = 1+0.3 cos(w_{A,1}t)$ and $a_2(t) = 0.2 + 0.1 cos(w_{A,2}t)$. 
See that $x_2(t)$ shows a linearly increasing frequency in time since $w_2(t) = w_{2,0} + w_{2,1}\frac{t}{2T}$. 
We assume that the signal generated has a time lapse of 10 seconds with sample frequency equal to 1000 observations per second.
In this particular example, we have chosen the following values $f_{1}=100Hz$, $f_{2,0}=10Hz$, $f_{2,1}=40Hz$, $f_{A,1}=1Hz$ and $f_{A,2}=5Hz$, being $w_.=2\pi f_.$. 
The simulated data are generated with the following \textsf{MATLAB} code:

\bigskip
\begin{ttfamily}
\begin{slshape}
\noindent
T = 10; \hspace{3cm} \% Time lapse, in seconds \newline
fs = 1000; \hspace{2.4cm} \% Sample frequency \newline
t = (1:T*fs)$'$; \newline
a1 = 1+0.3*cos(2*pi/1000*t); \newline
x1 = a1.*sin(2*pi/10*t); \newline
a2 = 0.2+0.1*cos(2*pi/200*t); \newline
w2 = 2*pi/100+2*pi/25*t/(2*T*fs); \newline
x2 = a2.*sin(w2.*t); \newline
x = x1+x2;
\end{slshape}
\end{ttfamily}
\bigskip

\begin{figure}[t!]
\centering
\includegraphics[width=0.9\linewidth]{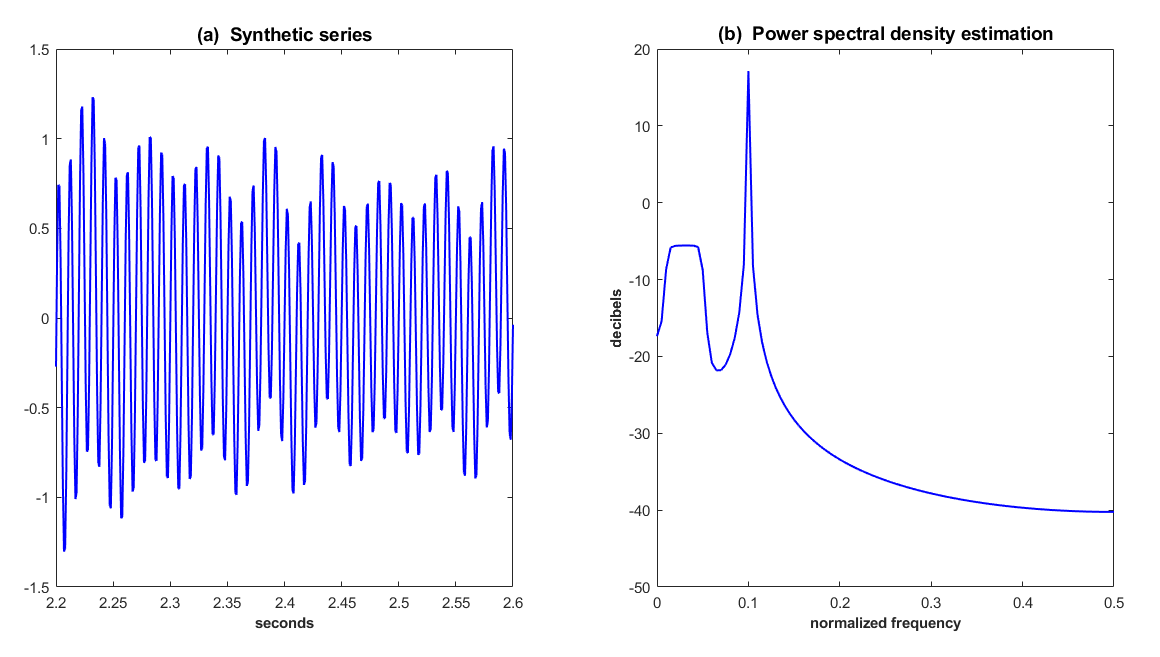}
\caption{Graph of the synthetic series (a) and its estimated power spectral density (b).}
\label{Synthetic_1} 
\end{figure}

The simulated signal is saved in \texttt{x}. The time lapse between 2.2 and 2.6 seconds is represented in the left panel of Figure \ref{Synthetic_1}.

To apply CiSSA we need to choose a value for $L$. It is advisable to choose a value of $L$ which is a multiple of the oscillation periods. 
In this case, the period associated with $x_1(t)$ is 10 and the periods associated with the frequency modulated signal $x_2(t)$ range from 20 to 100, and we have chosen $L=200$. 
The input parameter \texttt{H} (used to extend the data at the beginning and end of the sample) takes the value of 1 since this is an AM-FM signal. The code for signal extraction applying our proposal is:

\bigskip
\begin{ttfamily}
\begin{slshape}
\noindent
L = 200; \newline
H = 1; \hspace{3cm} \% Extension by Mirroring \newline
[Z, psd] = cissa(x,L,H);
\end{slshape}
\end{ttfamily}
\bigskip

The right hand side of Figure \ref{Synthetic_1} shows the estimated power spectral density which is the output saved on the variable \texttt{psd}\footnote{Notice that, due to symmetry, we only present normalized frequencies up to 0.5.}. 
From the latter, we can see two outstanding features. 
Firstly, a peak at the normalized frequency 0.1 that corresponds to $x_1(t)$. 
And, secondly, a plateau reflecting a constant gain between the normalized frequencies 0.01 and 0.05 associated with the modulated signal $x_2(t)$. 
Therefore, we should identify 2 sets in order to make the groups: 
the first contains the isolated normalized frequency linked to $x_1(t)$ that, according to (\ref{w_k}), corresponds to $k=21$; 
and, the second group includes the normalized frequencies of the plateau that characterizes $x_2(t)$ and are associated with the indexes $k=3,...,11$. 

The output $Z$ contains all the reconstructed signals by frequency, for $k=1,...,101$ and is one of the inputs of \texttt{group()}. The remaining two inputs are \texttt{psd} and \texttt{I}, where the latter contains the values of $k$ needed for grouping:

\bigskip
\begin{ttfamily}
\begin{slshape}
\noindent
I = cell(2); \hspace{0.2cm} \% 2 groups \newline
I\{1\} = 21; \hfill \% Component with the isolated frequency in the first group \newline
I\{2\} = (3:11); \% Components 3 to 11 in the second group \newline
[rc, sh, kg] = group(Z,psd,I);
\end{slshape}
\end{ttfamily}
\bigskip

The output has 3 outcomes: \texttt{rc} that is used to save the reconstructions of $x_1(t)$ and $x_2(t)$; the $2 \times 1$ vector \texttt{sh} that takes the values 95 and 4 indicating the percentage of variability explained by each of the two reconstructed components, respectively; and, finally, in this particular case, since the grouping has been introduced manually by the user \texttt{kg} = \texttt{I}. 

\begin{figure}[t!]
\centering
\includegraphics[width=0.9\linewidth]{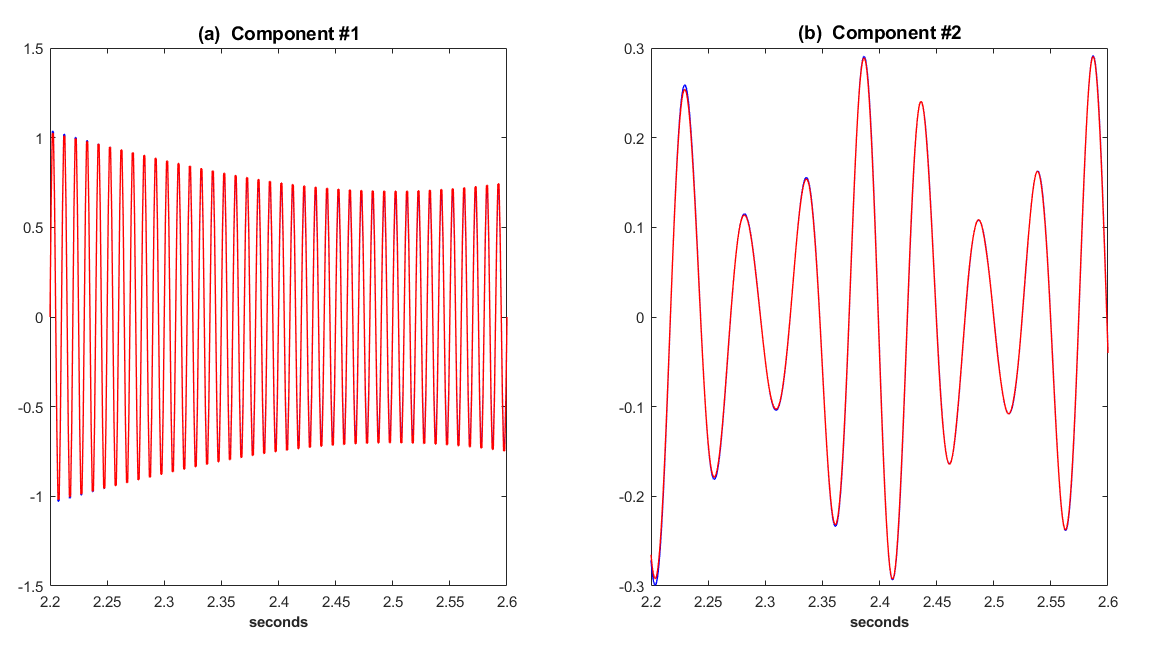}
\caption{Synthetic components (blue) and their corresponding CiSSA estimation (red).}
\label{Synthetic_2}
\end{figure}

Figure \ref{Synthetic_2} shows the reconstructed components stored in \texttt{rc} in red together with the components generated in blue for the time interval between 2.2 and 2.6 seconds in order to check the performance of the algorithm. 
As we can see, they are really close and CiSSA performs quite well.

\subsection{Speech Processing}

As a second example to illustrate the performance and use of options 2 to 4 of input \texttt{I} in \texttt{group()}, we downloaded a voiced speech from the Cambridge dictionary with the word “Hello”. 
The goal is to find a unique group able to synthesize the original word, that could be similar to de-noising.
Firstly, under the expert user option we decided on the groups after looking at the values of the estimated spectral power density. 
Secondly, under the semi-automated largest \texttt{psd} grouping, we form one group with the elementary reconstructed components with a frequency greater than a specified percentile of the power spectral density. 
And thirdly, as an alternative semi-automated criterion, the dimension reduction option returns one group that attains the desired share of explained variability. 

\begin{figure}[t!]
\centering
\includegraphics[width=0.9\linewidth]{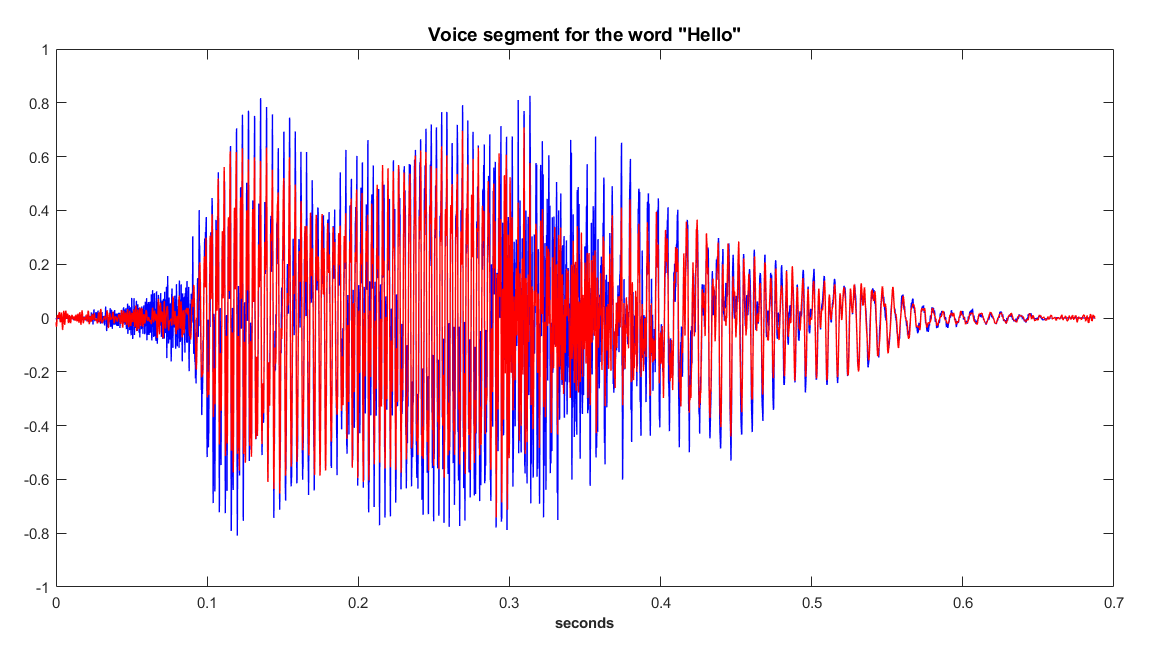}
\caption{The original voiced signal "Hello" (blue line) and its reconstruction (red line).}
\label{Hello_1}
\end{figure}

The voiced signal consists of 33000 observations and a sample frequency of 48000 Hz and is represented by the blue line in Figure \ref{Hello_1}. 

We ran \texttt{cissa()} with $L=6000$, for the window length\footnote{$L$ is either a multiple or a fraction of the sample frequency. In this case, we chose a fraction because the sample length is smaller than the sample frequency.} and used the default value for \texttt{H}, as the time series showed no stochastic trend and both options \texttt{H=0} or \texttt{H=1} are suitable options. 
The elementary reconstructed components by frequency output are stored in the $M$ columns of matrix \texttt{Z} and the power spectral density in the vector \texttt{psd}. 
In the three cases we are going to consider just one group and the differences that arise in the way to select it.

\begin{figure}[t!]
\centering
\includegraphics[width=0.9\linewidth]{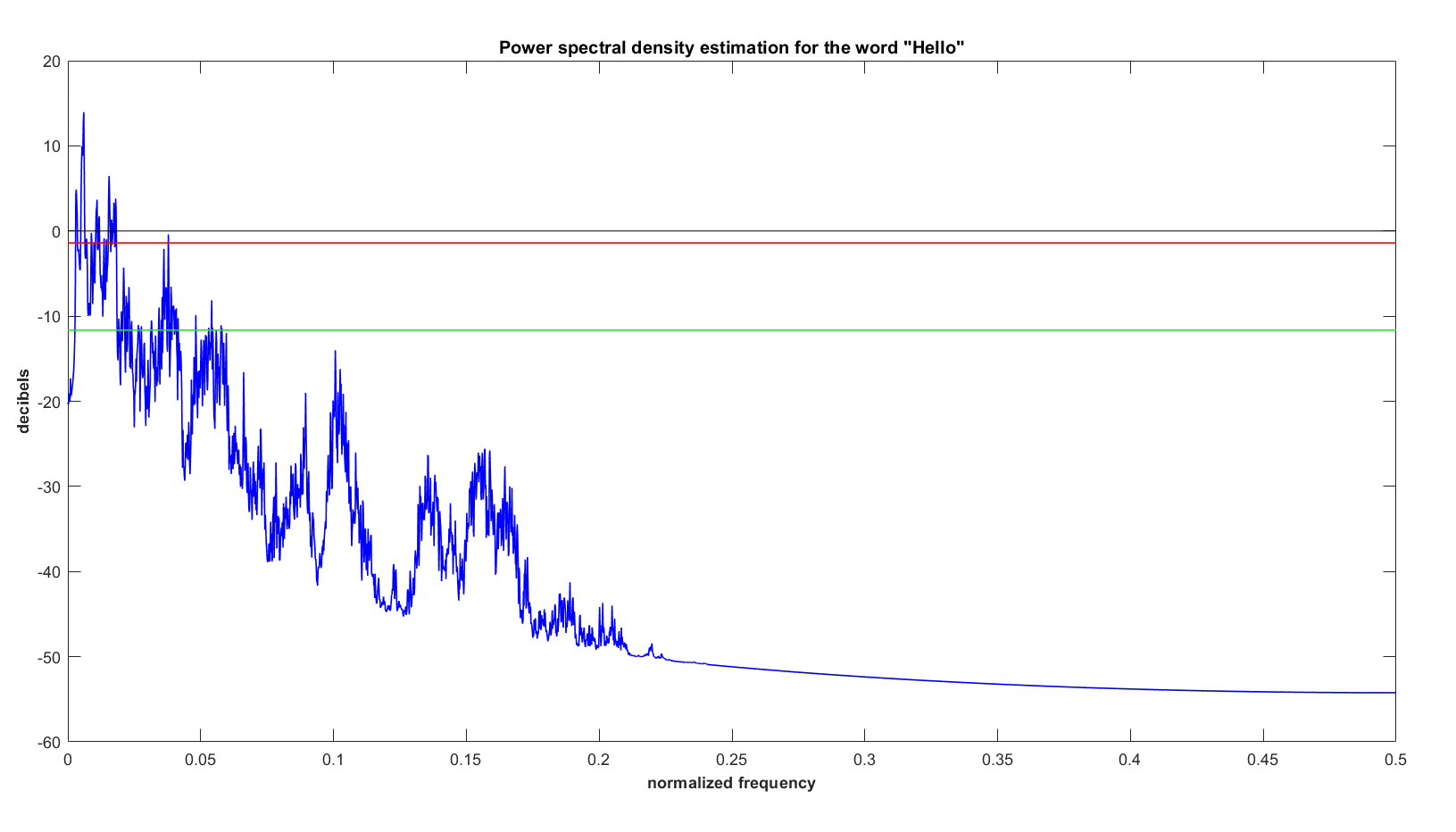}
\caption{Power spectral density in dB (blue line) of the voiced signal "Hello" and selected threshold (red line).}
\label{Hello_2}
\end{figure}

Firstly, in an expert user option, we selected the components and introduced them as a group in \texttt{group()}. 
We did so by looking at the values of the estimated power density (dB) in \texttt{psd}, that is represented in Figure \ref{Hello_2}. 
As a first attempt, we formed the group with the components with a positive value, that corresponded to normalized frequencies with \texttt{psd} over the black line in Figure \ref{Hello_2}. 
We stored the indexes $k$ of those components with positive power spectral density in \texttt{I} and used these values in \texttt{group()}.

\bigskip
\begin{ttfamily}
\begin{slshape}
\noindent
L = 6000; \newline
[Z, psd] = cissa(x, L); \newline
M = size(Z,2); \newline 
k = (1:M); \newline    
I = cell(1); \hspace{4.4cm} \% Define the number of groups (1 here) \newline
thres = 0; \hspace{4.9cm} \% Define a threshold for the psd \newline
I\{1\} = k(10*log10(psd(k))>thres); \hfill \% Select k with psd over the threshold \newline
[rc1, sh1, kg1] = group(Z,psd,I); 
\end{slshape}
\end{ttfamily}
\bigskip

The reconstructed signal is stored in \texttt{rc1}. In \texttt{sh1} we obtained the percentage of explained variability (74.3\%) in the first attempt. 
Finally, in \texttt{kg1} we saved the selected components that in this case coincided with \texttt{I}. 
The number of components with positive \texttt{psd} in dB was 29. As a result, the reconstructed signal with the previous grouping reproduces the sound /\textipa{@"l@U}/, leaving out the initial sound /\textipa{h}/. 
To incorporate the missing phoneme we change the threshold for selecting components. Choosing \texttt{thres=-1.4} solved the problem. 
We can identify that the $k$ linked to the sound /h/ is 228 that corresponds to the normalized frequency 0.0378.
Figure \ref{Hello_2} shows the estimated power spectral density (PSD) in dB with the above-mentioned threshold in red. The reconstructed signal (red line in Figure \ref{Hello_1}) quite accurately reproduces the sound /\textipa{h@"l@U}/.

In order to achieve the desired reconstruction in the previous example, we could alternatively introduce a group automatically formed by those components whose \texttt{psd} is over a percentile of the empirical distribution of the eigenvalues. 
To do so for the 95th percentile, select \texttt{I=-0.95} in \texttt{group()} as we show in the following piece of code:

\bigskip
\begin{ttfamily}
\begin{slshape}
\noindent
I = -0.95; \hspace{3cm} \% Share of explained variability \newline
[rc2, sh2, kg2] = group(Z,psd,I);
\end{slshape}
\end{ttfamily}
\bigskip

As in the previous case, \texttt{rc2} stores the reconstructed signal, \texttt{sh2} is now 96.4 \% and \texttt{kg2} contains the 150 indexes $k$ of the selected components. 

Thirdly and finally, we illustrate a different option for grouping using the percentage of accumulated explained variability. Choosing at least 80 \% of the explained variability, we introduced this input in the \texttt{group()} function as a negative number in order to distinguish it from the \texttt{psd} largest percentile option as the following piece of code illustrates:

\bigskip
\begin{ttfamily}
\begin{slshape}
\noindent
I = 0.80; \newline
[rc3, sh3, kg3] = group(Z,psd,I); 
\end{slshape}
\end{ttfamily}
\bigskip

Now \texttt{rc3} stores the reconstructed signal that is shown in red in Figure \ref{Hello_1}. The exact explained variability saved in \texttt{sh3} is 80.1 \% and \texttt{kg3} contains the 41 indexes $k$ of the selected components.
This solution also corresponds with the second selected threshold needed in order to incorporate the sound /h/ to reproduce the word "hello".

\subsection{Economic example}
Indicators of economic activity typically show regular patterns such as trend, seasonality and business cycle that are key to assessing the momentum of the economy.
In this respect, Statistical Offices regularly produce seasonally adjusted time series and analysts demand business cycle estimations of a myriad of economic indicators.
Given that for economists the frequencies of interest are known beforehand, CiSSA can be used to estimate these signals in an automated way.
In this section, we explain and illustrate the automated extraction of these signals by means of the use of \texttt{cissa()} and \texttt{group()} under the automated option.

\begin{figure}[t!]
\centering
\includegraphics[width=0.8\linewidth]{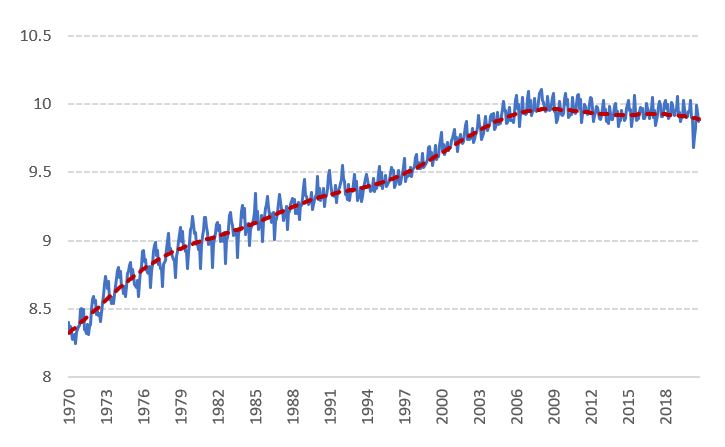}
\caption{Log Monthly Spanish Energy Consumption (in blue) and CiSSA estimated trend (dotted line in red).}
\label{OriginalTrend}
\end{figure}

The Spanish Energy Consumption, in millions of kilowatts, serves as a representative of time series of monthly economic activity. 
The source is REE (acronysm of Red Eléctrica de España, the sole transmission agent and operator of the Spanish electricity system). 
Figure \ref{OriginalTrend} shows the evolution of the log-transformation\footnote{SSA does not require log-transformation of the time series but, as typically done in business cycle analysis, we take logs in order to make the figures comparable with those usually handled by economic analysts.} of this indicator from January 1970 until October 2020. It clearly shows trend, business cycle oscillations and seasonality. 
The sample size is $T=610$ observations.

\begin{figure}[t!]
\centering
\includegraphics[width=0.8\linewidth]{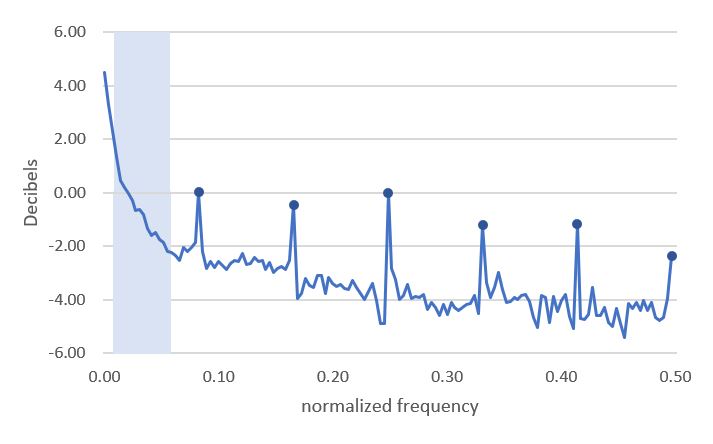}
\caption{Power spectral density estimation for the monthly Spanish Energy Consumption.}
\label{Psd}
\end{figure}

To perform \texttt{cissa()} we just needed to decide on the parameter $L$. 
In this case we have considered \texttt{L=288}. 
The main rationale for this selection for $L$ are that it has to be a number below $T/2$ and a multiple of the number of observations per year, and of the largest frequency considered (8 years, 96 months for business cycles)\footnote{Notice that using this reasoning, alternative possible values for $L$ could be $L=96$ or $L=192$. The main results are robust for the choice of $L$.}. 
\texttt{cissa()} output produced a matrix of $145$ elementary reconstructed components by frequency \texttt{Z} and a vector \texttt{psd} that stores the CiSSA estimated power spectral density at normalized frequencies given by (\ref{w_k}) , $k=1,2, ...,288$. 

The grouping of the reconstructed components by frequency to obtain the desired signals was carried out in an automated way running \texttt{group()} with \texttt{I=12} (the number of observations per year). 
Before analyzing the output produced, we highlighted how the groups are formed as fluctuations are assigned to each group according to their periodicity. 
It means that the trend is constructed as the sum of the components corresponding to periodicities greater than 8 years. Therefore, the group for the trend comprises $k=1$ ($\infty$) and $k=2, 3$ that, given (\ref{w_k}), will correspond to normalized frequencies below $1/288$ and $1/144$, i.e., 24 and 12 year cycles.
The business cycle is made of fluctuations between 1.5 and 8 years, therefore, it is the result of the sum of the components $k=4, ..., 17$. These correspond to normalized frequencies between 0.010 (1/96) and 0.056 (1/18) and are highlighted in grey in Figure \ref{Psd}.
Finally, seasonality gathers fluctuations within the year that correspond to normalized frequencies $1/12, 2/12, 3/12, 4/12, 5/12$ and $6/12$ and that correspond with the values of $k=25, 49, 73, 77, 121$ and $145$. The values of the power spectral density associated with the corresponding normalized frequencies for the selected values of $k$ are marked with a large dot in Figure \ref{Psd}.

The code for the previous analysis is very simple and is shown in the following lines:

\bigskip
\begin{ttfamily}
\begin{slshape}
\noindent
L = 288; \newline
[Z, psd] = cissa(log(x), L); \newline
I = 12; \hspace{3cm} \% Number of observations in the year \newline
[rc, sh, kg] = group(Z,psd,I);
\end{slshape}
\end{ttfamily}
\bigskip

\begin{figure}[t!]
\centering
\includegraphics[width=0.8\linewidth]{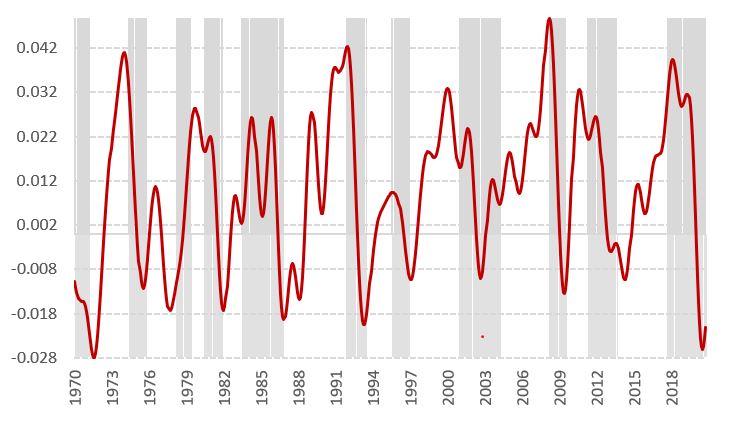}
\caption{Estimated Business Cycle for the monthly Spanish Energy Consumption and OECD announced recessions (shadowed areas).}
\label{BusinessCycle}
\end{figure}

\begin{figure}[t!]
\centering
\includegraphics[width=0.8\linewidth]{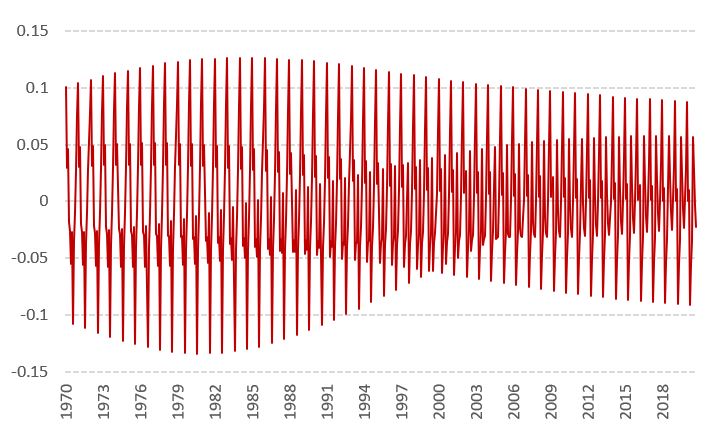}
\caption{Estimated seasonality for the monthly Spanish Energy Consumption.}
\label{Seasonality}
\end{figure}

The output of \texttt{group()} is composed of three items. 
Firstly, \texttt{rc} contains three time series that correspond to the trend, business cycle and seasonal component as they are plotted in Figures \ref{OriginalTrend}, \ref{BusinessCycle} and \ref{Seasonality}, respectively. 
Secondly, \texttt{sh} provides the share of the power spectral density for each group that are 88.2 \% for the trend, 8.5 \% for the business cycle and 1.5 \% for the seasonal component. 
Finally, \texttt{kg} stores the indexes of the selected components in each group.  

The analysis of the extracted signals reveals that the estimated trend, see the dotted red line in Figure \ref{OriginalTrend}, is able to reflect the long-term behaviour of the Spanish log of Energy Consumption. 
With regard to the business cycle, Figure \ref{BusinessCycle} shows that it matches the recessions as dated by the OECD for the Spanish economy.
Finally, Figure \ref{Seasonality} shows the changing seasonal behaviour of the Spanish energy consumption where the drops in summer are less pronounced in August, and have even become positive in June and July in recent times.

\section{Concluding remarks}
The \texttt{cissa()} and \texttt{group()} functions can be used to perform Circulant Singular Spectrum Analysis in a univariate time series. 
The only input needed are the data, the window length chosen for the analysis, and the option desired for grouping. 
The output gives the reconstructed components as well as the estimation of the power spectral density and the selected components for grouping. 
CiSSA can be applied in a variety of fields such as those described in the Applications section. 
The functions are useful for the analysis of stationary and non-stationary time series. 
They are also very useful for extracting components in a non-linear time series and due to its non-parametric nature, is not necessary to specify a model. 
Additionally, it can be run in an automated way when the frequency of oscillation is known beforehand, as is usually the case in economic analysis. 
The functions can be downloaded from the web page \url{https://es.mathworks.com/matlabcentral/fileexchange/84094-cissa-circulant-ssa-under-matlab?s_tid=srchtitle} or \url{https://github.com/jbogalo/CiSSA}.

\bibliography{References}

\end{document}